\documentstyle[aps,prd,multicol]{revtex}
\begin{document}
\draft
\title{ \hspace{5true in}
\begin{minipage}{2in}
\begin{flushright}
 {\rm UCSD/PTH/98-35} \\
 {\rm OHSTPY-HEP-T-98-016}
\end{flushright}
\end{minipage} \\ \vspace{5mm}
 Perturbative matching of lattice and continuum heavy-light\\
 currents with NRQCD heavy quarks}
\author{Colin~J.~Morningstar}
\address{Dept.~of Physics, University of California at San Diego,
  La Jolla, California 92093-0319}
\author{J.~Shigemitsu}
\address{Physics Department, The Ohio State University,
  Columbus, OH 43210}
\date{October 21, 1998}
\maketitle
\begin{abstract}
The temporal and spatial components of the heavy-light vector current
and the spatial components of the axial current are expressed in terms
of lattice-regulated operators suitable for simulations of $B$ and $D$
mesons.  The currents are constructed by matching the appropriate
scattering amplitudes in continuum QCD and a lattice model to
one-loop order in perturbation theory.  In the lattice theory,
the heavy quarks are treated using the nonrelativistic (NRQCD) formulation
and the light quarks are described by the tadpole-improved clover action.
The light quarks are treated as massless.  Our currents include
relativistic and discretization corrections through 
$O(\alpha_s/M, a \,\alpha_s)$, where $M$ is the heavy-quark mass,
$a$ is the lattice spacing, and $\alpha_s$ is the QCD coupling.  As in
our previous construction of the temporal component of the heavy-light
axial current, mixing between several lattice operators is encountered
at one-loop order, and $O(a \,\alpha_s)$ dimension-four improvement terms
are identified. 
\end{abstract}
\begin{multicols}{2}

\section{Introduction}

An important goal of lattice gauge theory is to provide estimates of the
hadronic matrix elements, such as those of the electroweak currents
and effective four-fermion operators, which are needed for precision tests of
the standard model.  Numerical simulations of quarks and gluons in
lattice-regulated QCD currently provide the only means of calculating
such matrix elements from first principles taking long-distance QCD
dynamics fully into account.  An important ingredient in these calculations
is the construction of the lattice current operators which match the
currents defined in the continuum to some desired accuracy.  In this paper,
we express the heavy-light vector and axial-vector currents in terms of
operators defined in a lattice model in which the heavy quarks are 
treated using the nonrelativistic (NRQCD) formulation\cite{cornell},
the light quarks are described by the tadpole-improved clover
action\cite{SWaction}, and the gluons are governed by the standard
Wilson action.  These heavy-light currents are important in
studies of heavy meson leptonic and semileptonic decays (for recent
reviews, see Refs.~\cite{review1,review2,review3}).  
The lattice currents are determined by matching scattering amplitudes
in the lattice model to those in continuum QCD to one-loop order
in perturbation theory.  The matching is carried out through 
$O(1/M,\alpha_s/M,a\,\alpha_s)$ where $M$ is the heavy quark mass,
$a$ is the lattice spacing, and $\alpha_s$ is the QCD coupling.
The light quarks are treated as massless.

Each heavy-light current $J_\mu$, defined in some continuum
renormalization scheme, can be written 
\begin{equation}  \label{jmatch}
  J_\mu \!=\! \sum_j C_j^J\ J_{\mu,{\rm lat}}^{(j)}
 +  O(\alpha_s^2,a^2\!\!,1/M_0^2, \alpha_s a/M_0),
\end{equation}
where $J_{\mu,{\rm lat}}^{(j)}$ are operators defined in the effective
lattice theory, $M_0$ is the heavy quark (bare) mass parameter appearing
in the lattice NRQCD action, and the $C_j^J$ coefficients are $c$-numbers
which depend only on $\alpha_s$ and $aM_0$. The goal is to identify the
operators $J_{\mu,{\rm lat}}^{(j)}$ and to calculate the matching
coefficients $C^J_j(\alpha_s,aM_0)$.  The procedure for doing this was
described in a previous article\cite{pert1} in which the temporal
component of the axial-vector current was studied.  For convenience,
we reiterate the salient steps: (1) select a quark-gluon scattering
process induced by the heavy-light current of interest and calculate the
one-loop amplitude for this process in continuum QCD; (2) expand the
amplitude in powers of $1/M$; (3) identify operators in the lattice theory,
usually by inspection, which reproduce the terms in this expansion;
(4) calculate the one-loop mixing matrix of these operators in the lattice
theory; and (5) adjust the $C^J_j$ coefficients to produce a linear
combination of lattice current operators whose one-loop scattering
amplitude agrees with that determined in step 2 to a given order
in $1/M$ and $a$.  In this paper, we apply the above procedure to
the spatial components $A_k$ of the axial-vector current and the 
spatial and temporal components of the vector current $V_\mu$.  We omit much
of the calculational details since they are similar to those already
described in Ref.~\cite{pert1}.

In Sec.~\ref{sec:opident}, we start from a continuum QCD calculation and
identify the current operators in the effective theory needed to reproduce
the continuum current through $O(\alpha_s/M)$.  In Sec.~\ref{sec:mix}, we
describe the lattice theory and the one-loop mixing calculation among the 
$J_{\mu,{\rm lat}}^{(j)}$.  The matching procedure is then completed
in Sec.~\ref{sec:match}.  In Sec.~\ref{sec:disc}, we discuss 
$O(a \, \alpha_s)$ corrections in the static limit, and issues pertaining
to terms that behave as $\alpha_s \, {\rm log}(aM)$ are dealt with in
Sec.~\ref{sec:largelog}.  The paper concludes with a summary of our
results in Sec.~\ref{sec:summary}.

\section{Operator identification}
\label{sec:opident}

The heavy-light vector and axial-vector currents are given as usual by 
$V_\mu(x) = \bar q(x)\,\gamma_\mu\,h(x)$ and 
$A_\mu(x) = \bar q(x)\,\gamma_5\gamma_\mu\,h(x)$, respectively, 
where $q(x)$ is the light quark field, $h(x)$ is the heavy quark field,
and $\gamma_\mu$ are the standard Dirac $\gamma$-matrices in
Euclidean space-time which satisfy
$\{\gamma_\mu , \gamma_\nu \} = 2 \, \delta_{\mu \nu}$, and
$\gamma_5 = \gamma_0 \gamma_1 \gamma_2\gamma_3$.
Euclidean-space four-vectors are defined in terms of Minkowski-space
four-vectors (indicated by an underline) using $ x_0 \!=\! i\,\underline{x}^0$
and $x_j \!=\! \underline{x}^j \!=\! - \underline{x}_j$, for $j\!=\!1,2,3$.
For the derivative operator, $\partial_0 \!=\! -i \underline{\partial}_0$ and
$\partial_j \!=\! \underline{\partial}_j \!=\! - \underline{\partial}^j$.
The currents are related by $ J_0 \!=\!  \underline{J}_0$, and 
$ J^j \!=\! -i \, \underline{J}^j$. For further details, see
Refs.~\cite{pert1,cjfrules}.

The first step in identifying the necessary operators in the effective
lattice theory is to calculate the matrix elements of the continuum QCD
currents for an incoming heavy quark of momentum $p$ and an outgoing
light quark of momentum $p^{\prime}$.  Using the on-shell mass and wave
function renormalization scheme in Feynman gauge and expanding in $1/M$
(except for $u_h(p)$), one finds to one-loop
order in perturbation theory
\begin{equation} \label{fqcdterms}
\langle \, q(p^{\prime}) \,| \,  J_{\mu} \, | \,h(p) \, {\rangle}_{QCD}
=  \bar{u}_q(p^{\prime})\  
 W^J_\mu(p^\prime,p)\ u_h(p),
\end{equation}
where
\begin{eqnarray} \label{fqcdtermsb}
W^J_\mu(p^\prime,p) &=&  a_1\ \Gamma^J_\mu 
 -  a_2 \ \frac{ip_\mu}{M}\;\Gamma^J_0\gamma_0 
 -  a_3\ \frac{p\!\cdot\!p^{\prime}}{M^2}\;\Gamma^J_\mu \nonumber \\ 
&& - \, a_4 \ \frac{ip^{\prime}_\mu}{M}\;\Gamma^J_0\gamma_0
  + a_5\ \frac{p\!\cdot\!p^{\prime}}{M^2}\frac{ip_\mu}{M}\;
 \Gamma^J_0\gamma_0   \nonumber \\ 
&& + \quad O(1/M^2),
\end{eqnarray}
with
\begin{eqnarray}
a_1  &=& \ 1\, + \,\frac{\alpha_s}{3 \pi} \left[ \;3\, \ln\frac{M}{\lambda}
\,-\, \frac{11}{4} \right],  \nonumber \\
a_2  &=& \qquad\ \frac{\alpha_s}{3 \pi} \ 2, \nonumber \\
a_3  &=& \qquad\  \frac{\alpha_s}{3 \pi} \left[ \;\;6\, \ln\frac{M}{\lambda}
\,-\,\frac{8\pi}{3} \,\frac{M}{\lambda}\,+ \,\frac{1}{2} \right],\nonumber\\
a_4  &=& \qquad\ \frac{\alpha_s}{3 \pi} \left[- 2\, \ln\frac{M}{\lambda}
\,+\, \frac{1}{2} \right],  \nonumber \\
a_5  &=& \qquad\ \frac{\alpha_s}{3 \pi} \left[ -4\, \ln\frac{M}{\lambda}
\,+\, 5 \right].
\end{eqnarray}
For the vector current, $\Gamma^V_\mu = \gamma_\mu$ and
for the axial-vector current, $\Gamma^A_\mu = \gamma_5 \gamma_\mu$.
$u_h(p)$ and $u_q(p^\prime)$ are the
standard spinors for the heavy and light quarks, respectively, which
satisfy the Dirac equation.  The light quark mass is set equal to zero.
Ultraviolet divergences are treated using dimensional regularization, and
fully anti-commuting $\gamma_5$ matrices are used.  We use a gluon
mass $\lambda$ to regulate infrared divergences.  Note that $M$ is the
heavy-quark pole mass.

In lattice NRQCD, the heavy quark is described in terms of a two-component
(in spin space) field $\psi(x)$.  The Dirac field $h(x)$ is related to 
$\psi(x)$ (and the antiquark field $\tilde{\psi}(x))$ by a unitary
Foldy-Wouthuysen transformation\cite{FWtrans},
\begin{equation} 
\label{fw}
 h(x) =  U_{FW}^{-1}
  \; \left(\begin{array}{c} \psi(x) \\ \tilde{\psi}(x) \end{array}\right).
\end{equation}
This transformation decouples the upper and lower components of the
Dirac field, thereby separating the quark field from the antiquark field.
To facilitate the identification of lattice NRQCD operators capable of
matching Eq.~(\ref{fqcdterms}), we similarly transform the external state
spinor $u_h(p)$ into a nonrelativistic Pauli spinor:
\begin{equation} \label{fwu}
  u_h(p) =  \left[ 1 - \frac{i}{2 M}\ \mbox{\boldmath$\gamma$}
   \!\cdot\! {\bf p} \right] \,u_Q(p)  +  O(1/M^2),
\end{equation}
where
\begin{equation}
 u_Q(p) = \left(\begin{array}{c} U_Q  \\ 0 \end{array} \right),
\end{equation}
and $U_Q$ is a two-component external state spinor depending
only on the spin of the heavy quark.  Note that we are working in the
Dirac-Pauli representation.  Using Eq.~(\ref{fwu}), the relation
$ \gamma_0 \, u_Q(p) = u_Q(p)$, and the Dirac equation for the light
quark $\bar u_q(p^\prime)p_0^\prime = -\bar u_q(p^\prime)
\mbox{\boldmath$\gamma$} \!\cdot\! {\bf p^\prime}\gamma_0$,
the spatial components of Eq.~(\ref{fqcdterms}) may be written
\begin{eqnarray} 
\label{fqcdtermsB}
\langle \, q(p^{\prime}) \,| \, J_k \, | \,h(p) \, {\rangle}_{QCD}
 &=&  \eta_0  \Omega_k^{(0)} + \eta_1 \Omega_k^{(1)}
 + \eta_2 \Omega_k^{(2)} \nonumber\\
&&+ \eta_3 \Omega_k^{(3)} + \eta_{4} \Omega_k^{(4)} \nonumber\\
&&  +  O(\alpha_s^2,1/M^2),
\end{eqnarray}
with
\begin{eqnarray}
\Omega_k^{(0)} & = &  \bar{u}_q(p^{\prime})\ \Gamma^J_k \ u_Q(p), 
\label{omega0}\\
\Omega_k^{(1)} & = & -  i\bar{u}_q(p^{\prime})\ \Gamma^J_k
  \frac{\mbox{\boldmath$\gamma$}
   \!\cdot\!{\bf p}}{2M}\ u_Q(p), \label{omega1}\\
\Omega_k^{(2)} & = &  i\bar{u}_q(p^{\prime})\ 
\frac{\mbox{\boldmath$\gamma$}
   \!\cdot\!{\bf p^\prime}}{2M}{\gamma}_0\ \Gamma^J_k\ u_Q(p), 
\label{omega2} \\
\Omega_k^{(3)} & = & -i \bar{u}_q(p^{\prime})\frac{p_k}{2M}\ \Gamma^J_0
\ u_Q(p), 
\label{omega3}\\
\Omega_k^{(4)} & = &  -i\bar{u}_q(p^{\prime})\ \frac{p^{\prime}_k}{2M}
 \ \Gamma^J_0\ u_Q(p).
\label{omega4}
\end{eqnarray}
The coefficients in Eq.~(\ref{fqcdtermsB}) are given by
\begin{eqnarray} \label{eta}
\eta_{0}  &=& \;\; a_1 \; = \; 1 + \alpha_s\ \tilde{B}_{0},  \nonumber \\
\eta_{1}  &=& \; \; \eta_{0}, \nonumber \\
\eta_{2}  &=&  2\,a_3 \, =  \alpha_s\ \tilde{B}_{2}, \nonumber \\
\eta_{3}  &=&  2\,a_2 \, =  \alpha_s\ \tilde{B}_{3}, \nonumber \\
\eta_{4}  &=&  2\,a_4 \, =  \alpha_s\ \tilde{B}_{4},
\end{eqnarray}
where 
\begin{eqnarray} \label{Bvalues}
\tilde{B}_{0}  &=& \frac{1}{3 \pi} \left[ 3\, \ln\frac{M}{\lambda}
- \frac{11}{4} \right],  \nonumber \\
\tilde{B}_{2}  &=& \frac{1}{3 \pi} \left[12\, \ln\frac{M}{\lambda}
 - \frac{16 \pi}{3} \frac{M}{\lambda} + 1\right], \nonumber \\
\tilde{B}_{3}  &=& \frac{1}{3 \pi} 4 , \nonumber \\
\tilde{B}_{4}  &=& \frac{1}{3 \pi} \left[-4 \, \ln\frac{M}{\lambda}
 + 1\right].
\end{eqnarray}

Having obtained the $1/M$ expansion of the above scattering amplitude
in continuum QCD, the next step is to identify operators in the lattice
effective theory that can reproduce the terms in this expansion.
An inspection of Eq.~(\ref{fqcdtermsB}) suggests immediately that matrix
elements of the following five lattice operators should be considered:
\begin{eqnarray}
 J^{(0)}_{k,{\rm lat}}(x) & = & \bar q(x) \,\Gamma^J_k\, Q(x),\label{Jop0}\\
 J^{(1)}_{k,{\rm lat}}(x) & = & \frac{-1}{2M_0} \bar q(x)
    \,\Gamma^J_k\,\mbox{\boldmath$\gamma\!\cdot\!\nabla$} \, Q(x),\label{Jop1}\\
 J^{(2)}_{k,{\rm lat}}(x) & = & \frac{-1}{2M_0} \bar q(x)
    \,\mbox{\boldmath$\gamma\!\cdot\!\overleftarrow{\nabla}$}
    \,\gamma_0\ \Gamma^J_k\, Q(x), \label{Jop2}  \\
 J^{(3)}_{k,{\rm lat}}(x) & = & \frac{-1}{2M_0} \bar q(x)\, \Gamma^J_0 \nabla_k 
\, Q(x)  \label{Jop3} \\
 J^{(4)}_{k,{\rm lat}}(x) & = & \frac{1}{2M_0} \bar q(x)
    \,\overleftarrow{\nabla}_k  \Gamma^J_0 \, Q(x), \label{Jop4} 
\end{eqnarray}
where $q(x)$ is now the light quark field in the lattice theory,
$M_0$ is the bare heavy quark mass, and $Q(x)$ is related to the heavy
quark field $\psi(x)$ in lattice NRQCD by
\begin{equation}
 Q(x) =  \left( \begin{array}{c} \psi(x) \\ 0 \end{array}\right).
\end{equation}
We use the bare quark mass $M_0$ in the above definitions since
it is the natural choice to use in lattice simulations and because
the pole mass $M$ is not well defined beyond perturbation theory.
Definitions of the lattice derivatives are given in the next section.

For the temporal components of the currents $J_0 = V_0$ or 
$ A_0$, the leading $1/M$ behavior is given by
\begin{eqnarray} 
\label{fqcdtermsBt}
\langle \, q(p^{\prime}) \,| \, J_0 \, | \,h(p) \, {\rangle}_{QCD}
 &=&  \eta^t_0\ \Omega_0^{(0)} + \eta^t_1 \ \Omega_0^{(1)}
 + \eta^t_2\ \Omega_0^{(2)} \nonumber\\
&&  +  O(\alpha_s^2,1/M^2),
\end{eqnarray}
with
\begin{eqnarray}
\Omega_0^{(0)} & = &  \bar{u}_q(p^{\prime})\ \Gamma^J_0\ u_Q(p), 
\label{omega0t}\\
\Omega_0^{(1)} & = & - i \bar{u}_q(p^{\prime})\ \Gamma^J_0
  \frac{\mbox{\boldmath$\gamma$}
   \!\cdot\!{\bf p}}{2M}\ u_Q(p), \label{omega1t}\\
\Omega_0^{(2)} & = & i\bar{u}_q(p^{\prime})\ \frac{\mbox{\boldmath$\gamma$}
   \!\cdot\!{\bf p^\prime}}{2M}\gamma_0\ \Gamma^J_0\ u_Q(p).
 \label{omega2t}
\end{eqnarray}
The coefficients in Eq.~(\ref{fqcdtermsBt}) are given by
\begin{eqnarray} \label{etat}
\eta^t_{0}  &=& \;\; (a_1 + a_2)\; = \; 1 + \alpha_s\ \tilde{B}^t_{0}, 
 \nonumber \\
\eta^t_{1}  &=& \;\;(a_1 - a_2) \;= \; 1 + \alpha_s\ \tilde{B}^t_{1},  
\nonumber \\
\eta^t_{2}  &=&  2\,(a_3 + a_4 + a_5 )\, =  \alpha_s\ \tilde{B}^t_{2},
\end{eqnarray}
where
\begin{eqnarray} \label{Bvaluest}
\tilde{B}^t_{0}  &=& \frac{1}{3 \pi} \left[ 3\, \ln\frac{M}{\lambda}
- \frac{3}{4} \right],  \nonumber \\
\tilde{B}^t_{1}  &=&  \frac{1}{3 \pi} \left[ 3\, \ln\frac{M}{\lambda}
- \frac{19}{4} \right],  \nonumber \\
\tilde{B}^t_{2}  &=& \frac{1}{3 \pi} \left[12 - \frac{16 \pi}{3}
 \frac{M}{\lambda} \right].
\end{eqnarray}
Again, it is straightforward to identify operators in the lattice 
effective theory which can reproduce these terms:
\begin{eqnarray}
 J^{(0)}_{0,{\rm lat}}(x) & = & \bar q(x) \,\Gamma^J_0\, Q(x),\label{Jop0t}\\
 J^{(1)}_{0,{\rm lat}}(x) & = & \frac{-1}{2M_0} \bar q(x)
    \,\Gamma^J_0\,\mbox{\boldmath$\gamma\!\cdot\!\nabla$} \, Q(x),\label{Jop1t}\\
 J^{(2)}_{0,{\rm lat}}(x) & = & \frac{-1}{2M_0} \bar q(x)
    \,\mbox{\boldmath$\gamma\!\cdot\!\overleftarrow{\nabla}$}
    \,\gamma_0\ \Gamma^J_0\, Q(x). \label{Jop2t}
\end{eqnarray}
These operators were previously used in the expansion of the temporal
component of the axial-vector heavy-light current 
in Refs.~\cite{pert1,pert2}.

\section{Mixing Matrix}
\label{sec:mix}

Having identified the necessary operators in the lattice-regulated
effective field theory, we next need to calculate the mixing matrix
$Z^{J}_{ij}$ of these operators defined by
\begin{eqnarray} 
\label{latterms}
\langle \, q(p^{\prime}) \,| \, J^{(i)}_{\mu,{\rm lat}}
  \, | \,h(p) \, {\rangle}_{\rm lat}
 &=&   \sum_j\ Z^{J}_{ij}\ \Omega_\mu^{(j)}  \nonumber\\
&& +   O(\alpha_s^2,1/M^2\!\!,a^2\!\!,\alpha_s a/M),
\end{eqnarray}
where $\Omega_\mu^{(j)}$ are the five $\Omega_k^{(j)}$ defined
in Eqs.~(\ref{omega0})-(\ref{omega4}) for $\mu=k=1,2,3$ and the 
three $\Omega_0^{(j)}$ defined in 
Eqs.~(\ref{omega0t})-(\ref{omega2t}) for $\mu=0$.  The determination of
the $Z^J_{ij}$ factors is the most difficult aspect of the current
construction.  These factors are determined numerically using lattice
perturbation theory.  The same methods used in Ref.~\cite{pert1}
are applied here. We also use the same lattice action as in
Ref.~\cite{pert1}, but for the convenience of the reader, we restate
these actions below.  The Feynman rules for our heavy and light quark
lattice action and further details of lattice perturbation theory are
given in Refs.~\cite{pert1,cjfrules}.  

For the heavy quark, the following NRQCD action 
density on the lattice\cite{cornell} is used:
 \begin{eqnarray} \label{nrqcdact}
  a{\cal L}_{NRQCD} &=&  \psi^\dagger(x)\ \psi(x)\nonumber\\
 &&- \psi^\dagger(x\!+\!a_t)\!
\left(1 \!-\!\frac{a \delta H}{2}\right)\!
 \left(1\!-\!\frac{aH_0}{2n}\right)^{n}\! 
 \frac{U^\dagger_4(x)}{u_0}\nonumber\\
&&\times \left(1\!-\!\frac{aH_0}{2n}\right)^{n}\!
\left(1\!-\!\frac{a\delta H}{2}\right)\! \psi(x),
 \end{eqnarray}
where 
 \begin{eqnarray}
 H_0 &=& - \frac{\Delta^{(2)}}{2M_0}, \label{Hkin} \\
 \delta H  &=& - c_B \,\frac{g}{2M_0}\,\mbox{\boldmath$\sigma$}
 \!\cdot\!{\bf B},\label{deltaH}
\end{eqnarray}
$U_\mu(x)$ are standard gauge-field link variables,
$n$ is an integer, $\sigma_j$ are the Pauli matrices,
$u_0$ is the mean link parameter introduced by the tadpole improvement
procedure\cite{viability}, and the QCD coupling $g$ is related to
$\alpha_s$ by $\alpha_s=g^2/(4\pi)$.
At tree level, $c_B = 1$; the one-loop contribution to $c_B$ is an
$O(\alpha^2)$ effect in our mixing matrix calculation and hence can be
ignored here.  The chromomagnetic field is given by 
$B_j(x)=-\frac{1}{2}\epsilon_{jlm}F_{lm}(x)$, where the Hermitian and
traceless field strength tensor $F_{\mu\nu}(x)$ is defined at the sites
of the lattice in terms of clover-leaf operators:
\begin{eqnarray}
 F_{\mu\nu}(x) &=& {\cal F}_{\mu\nu}(x) -
  {1\over 3}{\rm Tr}{\cal F}_{\mu\nu}(x), \nonumber\\
 {\cal F}_{\mu\nu}(x) &=& {-i\over 2a^2g}
\left( \Omega_{\mu\nu}(x)-\Omega^\dagger_{\mu\nu}(x)\right), \nonumber\\
\Omega_{\mu\nu}(x) &=&  {1\over 4u_0^4} \sum_{{\lbrace(\alpha,\beta)
   \rbrace}_{\mu\nu}}\!\!U_\alpha(x)U_\beta(x\!+\!a_\alpha)\nonumber\\
 &&\times \ U_{-\alpha}(x\!+\!a_\alpha\!+\!a_\beta)U_{-\beta}(x\!+\!a_\beta),
\label{fieldstrength}
\end{eqnarray}
with $\lbrace(\alpha,\beta)\rbrace_{\mu\nu} = \lbrace (\mu,\nu),
(\nu,-\mu), (-\mu,-\nu),(-\nu,\mu)\rbrace$ for $\mu\neq\nu$.
The lattice derivatives in our action and in the current operators 
are given by
\begin{eqnarray}
 a\nabla_\mu O(x) &=& \frac{1}{2u_0}[ U_\mu(x)O(x\!+\!a_\mu)\nonumber\\
 &&-U^\dagger_\mu(x\!-\!a_\mu) O(x\!-\!a_\mu)], \\
 O(x)\ a\!\overleftarrow{\nabla}_\mu &=& \frac{1}{2u_0}[ O(x\!+\!a_\mu)
  U^\dagger_\mu(x)\nonumber\\ &&-O(x\!-\!a_\mu) U_\mu(x\!-\!a_\mu) ], \\
 a^2\Delta^{(2)} O(x) &=& \sum_{k=1}^{3} (
 u_0^{-1}[ U_k(x)O(x\!+\!a_k)\nonumber\\
&&  +U^\dagger_k(x\!-\!a_k)
 O(x\!-\!a_k)]-2O(x) ), \\
 a^2\nabla^{(2)}O(x) &=&  \sum_{\mu=0}^{3} (
 u_0^{-1}[ U_\mu(x)O(x\!+\!a_\mu)\nonumber\\
 && +U^\dagger_\mu(x\!-\!a_\mu)
 O(x\!-\!a_\mu)] -2O(x) ),
\end{eqnarray}
where $O(x)$ is an operator defined at lattice site $x$ with appropriate
color structure. 

For the light quarks, we use the clover action\cite{SWaction},
\begin{eqnarray}
a{\cal L}_{light} &=&   \overline{q}\ \mbox{/\hspace{-0.6em}$\nabla$} q 
-  a\,\frac{r}{2} \,\overline{q}\, \nabla^{(2)} q
  + m_0\,\overline{q} q \nonumber\\
&&-iga \frac{r}{4} \sum_{\mu,\nu}\ \overline{q}\, \sigma_{\mu \nu} 
 F_{\mu \nu}\, q,
\label{cloveract}
\end{eqnarray}
where $\mbox{/\hspace{-0.6em}$\nabla$} = \sum_\mu \gamma_{\mu}
 \nabla_{\mu}$, $m_0$ is the bare light quark mass,
$\sigma_{\mu \nu} = \frac{1}{2} \left[\gamma_{\mu} , \gamma_{\nu} 
\right] $, and we set the Wilson parameter $r = 1$. 
The one-loop correction to the clover coefficient is an $O(\alpha_s^2)$
effect in our matching calculation and is neglected here.

At one-loop order, the mixing matrix elements may be written\cite{errcor}
\begin{equation} \label{zij}
Z^J_{ij} = \delta_{ij} \!+\! \alpha_s \left\{ \!\left[
  \textstyle{\frac{1}{2}} (\tilde{C}_q\! +\! \tilde{C}_Q) 
 \!+\! \tilde{C}_m (1\!-\!\delta_{i0})\right]\!\delta_{ij}
\! + \! \tilde{\zeta}^J_{ij} \right\},
\end{equation}
where $\tilde{C}_q$ and $\tilde{C}_Q$ are the contributions from the light-
and heavy-quark external leg corrections (that is, from wave function
renormalization factors), and $\tilde{\zeta}^J_{ij}$ denote
the contributions from the vertex corrections.  Our use of an on-shell
renormalization scheme with lattice operators defined in terms of the bare
mass $M_0$ is responsible for the term proportional to $\tilde{C}_m$, where
\begin{equation}
M =  [1 + \alpha_s\, \tilde{C}_m] \, M_0 + O(\alpha_s^2).
\label{massrenorm}
\end{equation}
Note that although the current operators $J_{\mu,{\rm lat}}^{(i)}$
are defined using $M_0$, the pole mass $M$ appears in $\Omega_\mu^{(j)}$.
Within the context of finite-order perturbation theory, the pole mass
is an observable quantity.  It is gauge invariant and regularization
scheme independent.  Hence, it is the natural heavy-quark mass
definition to use when matching amplitudes in different schemes.
However, in reality, a confined quark has no pole mass.  Once the
matching coefficients $C^J_j$ are obtained in terms of $aM$, it is then
necessary to eliminate any dependence on $M$ in favor of some quark
mass which is defined beyond perturbation theory.  Clearly, the mass
parameter $M_0$ appearing in the lattice action is the most suitable
and convenient choice.  The relationship between $M$ and $M_0$ given
in Eq.~(\ref{massrenorm}) is then used to re-express the $C^J_j$ coefficients
in terms of $M_0$.

The factors in Eq.~(\ref{zij}) may be further decomposed:
\begin{eqnarray}
 \tilde{C}_q &=& C_q + \frac{2}{3\pi}\ln a\lambda +C_q^{\rm TI},\nonumber\\
 \tilde{C}_Q &=& C_Q - \frac{4}{3\pi}\ln a\lambda,\nonumber\\
 \tilde{C}_m &=& C_m + C_m^{\rm TI},\nonumber\\
 \tilde{\zeta}^J_{ij} &=& \zeta^J_{ij} + \zeta^{J,{\rm TI}}_{ij}
 + \zeta^{J,{\rm IR}}_{ij}, \label{zijdecompose}
\end{eqnarray}
where $C_q$, $C_Q$, $C_m$, and $\zeta^J_{ij}$ are infrared finite and
independent of the tadpole improvement factor $u_0$, and
$\zeta^{J,{\rm IR}}_{ij}$ and $\zeta^{J,{\rm TI}}_{ij}$
contain the infrared divergences and tadpole improvement contributions,
respectively, from the vertex corrections.  Contributions to $\tilde{C}_q$
and $\tilde{C}_m$ from the tadpole improvement counterterms are
denoted by $C_q^{\rm TI}$ and $C_m^{\rm TI}$, respectively.
The tadpole improvement terms are
\begin{eqnarray}
 C_q^{\rm TI} &=& -u_0^{(2)},\nonumber\\
 C_m^{\rm TI} &=& -u_0^{(2)} \, \left( 1 
 - \frac{3}{2\, n\, aM_0}\right), \nonumber\\
 \zeta^{J,{\rm TI}}_{ij} &=& u_0^{(2)}\delta_{ij}(1-\delta_{i0}),
\end{eqnarray}
where $u_0=1-\alpha_s u_0^{(2)}+O(\alpha_s^2)$.
For the usual plaquette definition $u_0=\langle \frac{1}{3}{\rm Tr}
U_\Box\rangle^{1/4}$ in the Wilson gluonic action, $u_0^{(2)}=\pi/3$.
For massless clover quarks, $C_q = 1.030$.  Values for $C_Q$ and $C_m$
are given in Ref.~\cite{pert1}. 

\section{Matching}
\label{sec:match}

To complete the operator construction for $J_k$, we invert
Eq.~(\ref{latterms}) to get $\Omega_\mu^{(j)} = \sum_i(Z^J)^{-1}_{ji} 
\langle J^{(i)}_{\mu,{\rm lat}} \rangle_{{\rm lat}}$, substitute
this into Eq.~(\ref{fqcdtermsB}), peel off the dependence on the
external states, and use Eqs.~(\ref{eta}), (\ref{Bvalues}), (\ref{zij}),
(\ref{massrenorm}), and (\ref{zijdecompose}) to obtain
\begin{eqnarray} \label{ajalph2}
 J_k & = & \ \ \; \left( 1 + \alpha_s \left[ B_{0} 
 - C_{Qq}
 - \tau^{J_k}_0 \; \right]\right)
 \ J_{k,{\rm lat}}^{(0)}  \nonumber \\
  & & + \left( 1 + \alpha_s \left[ B_{1} 
 - C_{Qqm} -\tau^{J_k}_1 - \tau_1^{\rm TI}\right]\right)
 \ J_{k,{\rm lat}}^{(1)} \nonumber \\
& & + \qquad \;\;  \alpha_s \left[ B_{2} - \tau^{J_k}_2  \;\right]
 \ J_{k,{\rm lat}}^{(2)}  \nonumber \\
& & + \qquad \;\;  \alpha_s \left[ B_{3} - \tau^{J_k}_3  \;\right]
 \ J_{k,{\rm lat}}^{(3)}  \nonumber \\
& & + \qquad \;\;  \alpha_s \left[ B_{4} - \tau^{J_k}_4  \;\right]
 \ J_{k,{\rm lat}}^{(4)}   \nonumber\\
&& + \quad  O(\alpha_s^2,a^2,1/M^2,\alpha_s a/M),
\end{eqnarray}
where $\tau^{J_k}_j = \zeta^{J_k}_{0j} + \zeta^{J_k}_{1j}$,
$\tau_1^{\rm TI} = u_0^{(2)}$,
\begin{eqnarray} \label{bk}
B_{0} &=& \frac{1}{\pi}\left[\ln(aM_0)- \frac{11}{12}\right] = B_{1},\nonumber \\
B_{2} &=& \frac{1}{\pi}\left[4\,\ln(aM_0) + \frac{1}{3}\right], \nonumber \\
B_{3} &=& \frac{4}{3\pi}, \nonumber \\
B_{4} &=& \frac{1}{\pi}\left[-\frac{4}{3} \ln(aM_0) + \frac{1}{3}\right],
\end{eqnarray}
and
\begin{eqnarray}
C_{Qq} &=&\frac{1}{2}(C_q+C_q^{\rm TI}+ C_Q),\\
C_{Qqm} &=& C_{Qq} +C_m+C_m^{\rm TI}.
\end{eqnarray}
As expected, the infrared divergences from the various terms in the
expansion coefficients cancel.  Results for $\tau^{V_k}_j$ and
$\tau^{A_k}_j$ for various values of $aM_0$ are listed in 
Tables~\ref{tab:one} and \ref{tab:two}, respectively.

For the temporal component of the vector current, we substitute the
inverted Eq.~(\ref{latterms}) into Eq.~(\ref{fqcdtermsBt}), remove the
dependence on the external states, and use Eqs.~(\ref{etat}),
(\ref{Bvaluest}), (\ref{zij}), (\ref{massrenorm}), and (\ref{zijdecompose})
to obtain
\begin{eqnarray} \label{ajalphv0}
 V_0 & = & \ \ \; \left( 1 + \alpha_s \left[ B^t_{0} 
 - C_{Qq} - \tau^{V_0}_0 \; \right]\right)
 \ J_{0,{\rm lat}}^{(0)}  \nonumber \\
  & & + \left( 1 + \alpha_s \left[ B^t_{1} 
 - C_{Qqm} -\tau^{V_0}_1 - \tau_1^{\rm TI}\right]\right)
 \ J_{0,{\rm lat}}^{(1)} \nonumber \\
& & + \qquad \;\;  \alpha_s \left[ B^t_{2} - \tau^{V_0}_2  \;\right]
 \ J_{0,{\rm lat}}^{(2)} \nonumber\\
&& + \quad O(\alpha_s^2,a^2,1/M^2,\alpha_s a/M),
\end{eqnarray}
where
\begin{eqnarray} \label{bt}
B^t_{0} &=& \frac{1}{\pi}\left[\ln(aM_0) - \frac{1}{4}\right],  \nonumber \\
B^t_{1} &=& \frac{1}{\pi}\left[\ln(aM_0) - \frac{19}{12}\right], \nonumber \\
B^t_{2} &=& \frac{4}{\pi}.
\end{eqnarray}
Again, all infrared divergences cancel.  Results for $\tau^{V_0}_j$
for various $aM_0$ values are given in Table~\ref{tab:three}.

From Eqs.~(\ref{ajalph2}) and (\ref{ajalphv0}), one easily obtains the
matching coefficients $C^J_j$ of Eq.~(\ref{jmatch}) which expresses
the heavy-light currents in terms of operators suitable for numerical
simulations using lattice NRQCD.  These coefficients can be written
in the form
\begin{equation}
 C_i^J(\alpha_s,aM_0)  = \left\{
 \begin{array}{r@{\hspace{2ex}}l}
 1+\alpha_s\ \rho^J_i(aM_0) +O(\alpha_s^2), &  (i=0,1), \\[5pt]
  \alpha_s\ \rho^J_i(aM_0)+O(\alpha_s^2), &  (i\geq 2). \end{array} \right.
\label{cjdef}
\end{equation}
The values for $\rho_i^{V_k}$, $\rho_i^{A_k}$, and $\rho_i^{V_0}$
for various values of $aM_0$ are listed in Tables~\ref{tab:four},
\ref{tab:five}, and \ref{tab:six}, respectively.

\section{ $O(\lowercase{a} \alpha_s)$ Corrections}
\label{sec:disc}

In this section, discretization corrections in the static limit
$aM_0\rightarrow\infty$ are discussed.  To study such corrections,
we calculated all $\tau_i^J$ for several large values of
the heavy quark mass.  With the exception of $\tau_2^J$, all of
the $\tau_i^J$ tended tamely to finite values in the static limit.
However, as $aM_0$ became large, we found that the magnitude of $\tau_2^J$
began to grow linearly with $aM_0$:
\begin{equation}
\tau^J_2 \stackrel{aM_0\gg 1}{\longrightarrow} 
 -\,2a M_0\ \zeta^J_{disc} + {\rm constant},
\end{equation}
where the $\zeta^J_{disc}$ are constants independent of $aM_0$.
We determined the $\zeta^J_{disc}$ by computing $\tau_2^J$ for
$aM_0=10.0,25.0,100.0,400.0,1000.0$ and $5000.0$.  We then
fit the results for $-\tau_2^J/(2aM_0)$ to a quadratic polynomial
in $1/(aM_0)$ and obtained the following limiting values:
\begin{eqnarray}
\zeta^{A_0}_{disc} &=& \hspace{1.8ex} 1.029(2),\\
\zeta^{V_k}_{disc} &=& \hspace{1.8ex} 1.031(1),\\
\zeta^{A_k}_{disc} &=& -0.063(1),\\
\zeta^{V_0}_{disc} &=& -0.063(1).
\end{eqnarray}
The discretization factor $\zeta^{A_0}_{disc}$ for the temporal
component of the axial-vector current was previously calculated
in Ref.~\cite{pert1}, but its determination was not done very accurately.
We have recalculated $\zeta^{A_0}_{disc}$ here to a much higher
precision.
  
Since the contributions proportional to $\zeta^J_{disc}$ are purely
$O(a\alpha_s)$ discretization corrections, one may absorb these terms
into the lattice current operators.  Improved current operators can be
defined using
\begin{eqnarray}
{\hat J}^{(0)}_{\mu,{\rm lat}}(x) &=& J^{(0)}_{\mu,{\rm lat}}(x) 
+ \alpha_s \ \zeta^J_{disc}\ J^{(disc)}_{\mu,{\rm lat}}(x), \label{Jimp} \\
{\hat J}^{(j)}_{\mu,{\rm lat}}(x) &=& J^{(j)}_{\mu,{\rm lat}}(x),
 \quad (j>0),
\end{eqnarray}
where
\begin{eqnarray}
 J^{(disc)}_{\mu,{\rm lat}}(x) & = & 2 aM_0 \ J^{(2)}_{\mu,{\rm lat}}(x),
\nonumber\\
&=&-a\, \bar q(x)
    \,\mbox{\boldmath$\gamma\!\cdot\!\overleftarrow{\nabla}$}
    \,\gamma_0\Gamma^J_\mu\, Q(x). \label{Jdisc}
\end{eqnarray}
The mixing of the improved operators is given by
\begin{eqnarray} 
\label{lattermsd}
\langle \, q(p^{\prime}) \,| \, {\hat J}^{(i)}_{\mu,{\rm lat}}
  \, | \,h(p) \, {\rangle}_{\rm lat}
 &=&   \sum_j\ {\hat Z}^{J}_{ij}\ \Omega_\mu^{(j)}  \nonumber\\
&& +   O(\alpha_s^2,1/M^2\!\!,a^2\!\!,\alpha_s a/M),
\end{eqnarray}
where
\begin{eqnarray}
{\hat Z}^J_{0i} &=& Z^J_{0i} + 2\alpha_s\,aM_0\zeta^J_{disc}\ \delta_{2i},\\
{\hat Z}^J_{ji} &=& Z^J_{ji}, \quad (j>0).
\end{eqnarray}
Then the coefficients ${\hat C}^J_j$ of these operators are
\begin{eqnarray}
 {\hat C}^J_2 &=& C^J_2 -2aM_0\ \zeta^J_{disc}\ \alpha_s,\\
 {\hat C}^J_i &=& C^J_i, \quad (i\neq 2),
\end{eqnarray}
and the heavy-light currents are given by
\begin{equation}  \label{jimpmatch}
  J_\mu \!=\! \sum_j {\hat C}_j^J\ {\hat J}_{\mu,{\rm lat}}^{(j)}
 +  O(\alpha_s^2,a^2\!\!,1/M_0^2, \alpha_s a/M_0).
\end{equation}
Terms which grow linearly with $aM_0$ are no longer present in
${\hat C}_j^J$ and ${\hat Z}^J_{ij}$.
Note that in the lattice NRQCD approach, discretization and relativistic
corrections are intertwined since $O(a)$ and $O(1/M)$ interactions are
treated as equally important.

\section{Large Logarithms}
\label{sec:largelog}

Another feature of the ${\hat C}^J_j$ matching coefficients is the presence of 
$\ln(aM_0)$ terms.  The simulations of heavy-light systems carried out to
date have used bare heavy quark mass values in the range 
$ aM_0 \sim 1.6\; - \; 4.0 $ where these logarithms are not large. 
However, in the large $aM_0$ limit, these logarithms must be treated
with care.  The logarithms appearing in the coefficients ${\hat C}_j^J$ for
$j>0$ are tamed by the $1/(aM_0)$ factors in their corresponding current
operators.  However, the logarithm appearing in the ${\hat C}_0^J$ coefficient
becomes problematical and must be dealt with using the renormalization
group.

In matching calculations between QCD and various continuum effective theories, 
one usually encounters similar logarithms of the form $\ln(M/\mu)$,
where $\mu$ is some scale introduced by the renormalization procedure.
Such logarithms are summed using the renormalization group (RG) equations
which follow from the requirement that physical quantities must not depend
on $\mu$.  Since $\mu$ appears only inside the logarithms, one ends up with
simple anomalous-dimension matrices and RG equations which can be solved
straightforwardly.  The situation is more complicated in the present
calculation.  The role of $\mu$ is taken over by the inverse lattice
spacing $1/a$, and $a$ appears not only in the logarithms, but also in
other places, such as the $\tau_j^J$ which are complicated functions of 
$aM_0$.  Furthermore, the ultraviolet cutoff $1/a$ is an integral part of
our effective theory and we cannot take $a \rightarrow 0$.  

As discussed in Ref.~\cite{pert1}, the observation that the left-hand side
of Eq.~(\ref{jmatch}) is independent of the lattice spacing can be
exploited to derive an RG equation for the ${\hat C}^J_j$ coefficients.
This equation describes the change in the ${\hat C}^J_j$ as the lattice spacing
is varied.  Collecting the ${\hat C}^J_j$ coefficients into a vector, the
RG equation may be written
\begin{equation}
 \left(a\frac{d}{da}+(\gamma^{J})^{\rm tr}\right) \vec{{\hat C}^J} = 0,
\label{rg1}
\end{equation}
where the anomalous dimension matrix is given by
\begin{equation} \label{gamma}
\gamma^J_{ij}(\alpha_s,aM_0) = \sum_k
 \left(a\frac{d}{da}{\hat Z}^J_{ik}\right) ({\hat Z}^J)^{-1}_{kj}.
\end{equation}
The right-hand side of Eq.~(\ref{gamma}) is a complicated function of $aM_0$.
Once $\gamma^J_{ij}$ is determined numerically for a large range of $aM_0$
values, Eq.~(\ref{rg1}) can be solved by numerical methods.  Our primary
concern was the determination of the matching coefficients ${\hat C}^J_j$
for values of $aM_0$ relevant for simulations of $B$ and $D$ mesons, and for
such values, RG improvement was not needed.  Hence, we have not attempted to
obtain the entire anomalous dimension matrices for our current operators.

\section{Summary}
\label{sec:summary}

In summary, the spatial components of the heavy-light axial-vector
current $A_k$ and all components of the heavy-light vector current $V_\mu$
were expressed in terms of lattice operators suitable for use in simulations
of $B$ and $D$ mesons.  In the lattice theory, the heavy quarks were treated
using the NRQCD formulation, the light quarks were described by the
tadpole-improved clover action, and the standard Wilson action was used
for the gluons.  The light quarks were treated as massless.
The expansions were carried out to $O(1/M)$ by matching appropriate
scattering amplitudes to one-loop order in perturbation theory.
We found
\begin{equation}
  J_\mu \!=\! \sum_{j=0}^{N_{J_\mu}\!-\!1} C_j^J\ J_{\mu,{\rm lat}}^{(j)}
 +  O(\alpha_s^2,a^2\!\!,1/M_0^2, \alpha_s a/M_0),
\end{equation}
where $N_{A_k}=N_{V_k}=5$ and $N_{V_0}=3$.  The lattice current operators
are given by
\begin{eqnarray}
 J^{(0)}_{\mu,{\rm lat}}(x) & = & \bar q(x) \,\Gamma^J_\mu\, Q(x),\\
 J^{(1)}_{\mu,{\rm lat}}(x) & = & \frac{-1}{2M_0} \bar q(x)
    \,\Gamma^J_\mu\,\mbox{\boldmath$\gamma\!\cdot\!\nabla$} \, Q(x),\\
 J^{(2)}_{\mu,{\rm lat}}(x) & = & \frac{-1}{2M_0} \bar q(x)
    \,\mbox{\boldmath$\gamma\!\cdot\!\overleftarrow{\nabla}$}
    \,\gamma_0\ \Gamma^J_\mu\, Q(x),   \\
 J^{(3)}_{k,{\rm lat}}(x) & = & \frac{-1}{2M_0} \bar q(x)
    \, \Gamma^J_0 \nabla_k \, Q(x)  \\
 J^{(4)}_{k,{\rm lat}}(x) & = & \frac{1}{2M_0} \bar q(x)
    \,\overleftarrow{\nabla}_k  \Gamma^J_0 \, Q(x), 
\end{eqnarray}
with $\Gamma^V_\mu = \gamma_\mu$ and $\Gamma^A_\mu 
= \gamma_5 \gamma_\mu$, and the coefficients are
\begin{equation}
 C_i^J  = \left\{
 \begin{array}{r@{\hspace{1ex}}l}
 1+\alpha_s\ \rho^J_i(aM_0) +O(\alpha_s^2), &  (i=0,1), \\[5pt]
  \alpha_s\ \rho^J_i(aM_0)+O(\alpha_s^2), &  (i\geq 2), \end{array} \right.
\end{equation}
where values of the $\rho^J_i(aM_0)$ are listed in
Tables~\ref{tab:four}-\ref{tab:six}.  The currents can also be
expressed in terms of improved current operators 
${\hat J}^{(j)}_{\mu,{\rm lat}}(x)$ as shown in Eq.~(\ref{jimpmatch}).

This completes the matching calculation through $O(\alpha_s/M)$ and 
$O(a \, \alpha_s)$ for all components of the vector and axial-vector 
heavy-light currents.  Our matching coefficients have already 
been applied in leptonic $B$ and $B^*$ meson decays to extract the
$f_{PS}$ and $f_V$ decay constants\cite{fb1,fb2,sara,hein}. 
They are also relevant for studies of $B  \rightarrow \pi$ or
$\rho$ semileptonic decays.  In this article, we presented results only
for the simple NRQCD action of Eqs.~(\ref{nrqcdact})-(\ref{deltaH}).
Matching coefficients for other NRQCD actions which have appeared in the
literature and for different values of $(aM_0, n)$ are also available.
For example, Ref.~\cite{fb1} used a slightly different action, and
an action with higher-order improvement terms was employed in 
Refs.~\cite{fb2,hein}.  In all cases, we find that in the useful range
$1.0 \leq aM_0 \leq 10.0$, the one-loop coefficients exhibit only a mild
dependence on $aM_0$ and do not become particularly large.

\section*{Acknowledgments}
This work was supported by the U.S.~DOE, Grants No.~DE-FG03-97ER40546
and DE-FG02-91ER40690, and by NATO grant CRG 941259.


\end{multicols}


\vspace{2cm}
\begin{table}
\begin{center}
\begin{minipage}{5 true in}
\caption[tabone]{Values of the coefficients $\tau^{V_k}_i$ corresponding to
  the spatial components of the vector current $V_k$
  for various values of the bare heavy-quark mass $aM_0$ and NRQCD
  stability parameter $n$.  Uncertainties in the determinations of these
  parameters due to the use of Monte Carlo integration are included.
\label{tab:one}}
\begin{center}
\begin{tabular}{rrrrrrr}
\multicolumn{1}{c}{$aM_0$} &
\multicolumn{1}{c}{$n$} &
\multicolumn{1}{c}{$\tau^{V_k}_0$} &
\multicolumn{1}{c}{$\tau^{V_k}_1$} &
\multicolumn{1}{c}{$\tau^{V_k}_2$} &
\multicolumn{1}{c}{$\tau^{V_k}_3$} &
\multicolumn{1}{c}{$\tau^{V_k}_4$} \\ \hline 
10.0 & 1 & 0.8189(1) &$-0.588(3) $ &$-13.93(1)\hspace{1.2ex} $ &$-0.566(4)$  &
 $-0.940(5)$ \\
 7.0 & 1 & 0.7880(1) &$-0.640(2) $ & $-8.150(7)$ &$-0.508(3)$  &
 $-0.909(3)$ \\
 4.0 & 1 & 0.7221(1) &$-0.757(5) $ & $-2.809(7)$ &$-0.385(7)$  &
 $-0.836(7)$ \\
 4.0 & 2 & 0.7340(1) &$-0.742(4) $ & $-3.084(8)$ &$-0.422(6)$  &
 $-0.842(7)$ \\
 3.5 & 2 & 0.7176(2) &$-0.775(4) $ & $-2.310(5)$ &$-0.392(6)$  &
 $-0.817(6)$ \\
 3.0 & 2 & 0.6982(1) &$-0.821(4) $ & $-1.594(4)$ &$-0.351(5)$  &
 $-0.791(5)$ \\
 2.7 & 2 & 0.6840(2) &$-0.856(3) $ & $-1.196(4)$ &$-0.323(5)$  &
 $-0.767(5)$ \\
 2.5 & 2 & 0.6732(2) &$-0.877(3) $ & $-0.955(4)$ &$-0.309(5)$  &
 $-0.750(4)$ \\
 2.0 & 2 & 0.6410(1) &$-0.962(3) $ & $-0.427(3)$ &$-0.254(4)$  &
 $-0.693(3)$ \\
 1.7 & 2 & 0.6170(2) &$-1.027(2) $ & $-0.188(2)$ &$-0.218(3)$  &
 $-0.647(3)$ \\
 1.6 & 2 & 0.6075(2) &$-1.051(2) $ & $-0.123(2)$ &$-0.205(3)$  &
 $-0.627(2)$ \\
 1.2 & 3 & 0.5768(2) &$-1.175(2) $ & $-0.081(2)$ &$-0.150(3)$  &
 $-0.534(2)$ \\
 1.0 & 4 & 0.5559(2) &$-1.265(2) $ & $-0.148(1)$ &$-0.108(3)$  &
 $-0.464(2)$ \\
 0.8 & 5 & 0.5249(2) &$-1.387(2) $ & $-0.288(1)$ &$-0.049(3)$  &
 $-0.363(1)$  \\
\end{tabular}
\end{center}
\end{minipage}
\end{center}
\end{table}

\begin{table}
\begin{center}
\begin{minipage}{5 true in}
\caption[tabtwo]{Values of the coefficients $\tau^{A_k}_i$ corresponding to
  the spatial components of the axial-vector current $A_k$,
  similar to Table~\protect\ref{tab:one}.
\label{tab:two}}
\begin{center}
\begin{tabular}{rrrrrrr}
\multicolumn{1}{c}{$aM_0$} &
\multicolumn{1}{c}{$n$} &
\multicolumn{1}{c}{$\tau^{A_k}_0$} &
\multicolumn{1}{c}{$\tau^{A_k}_1$} &
\multicolumn{1}{c}{$\tau^{A_k}_2$} &
\multicolumn{1}{c}{$\tau^{A_k}_3$} &
\multicolumn{1}{c}{$\tau^{A_k}_4$} \\ \hline
10.0 & 1 & 0.2634(1) &$-1.098(4)$ &2.624(18)\hspace{-1.2ex} &0.611(6) &$-0.7291(2)$ \\
 7.0 & 1 & 0.2685(1) &$-1.145(3)$ &2.137(12)\hspace{-1.2ex} &0.636(4) &$-0.7017(2)$ \\
 4.0 & 1 & 0.2817(1) &$-1.254(2)$ &1.494(6)  &0.691(2) &$-0.6401(2)$ \\
 4.0 & 2 & 0.2742(1) &$-1.197(2)$ &1.475(6)  &0.590(2) &$-0.6407(2)$ \\
 3.5 & 2 & 0.2766(1) &$-1.225(2)$ &1.339(5) &0.595(2) &$-0.6219(2)$ \\
 3.0 & 2 & 0.2793(1) &$-1.259(1)$ &1.168(4) &0.598(2) &$-0.5977(2)$ \\
 2.7 & 2 & 0.2812(1) &$-1.284(1)$ &1.065(3) &0.598(2) &$-0.5796(2)$ \\
 2.5 & 2 & 0.2824(1) &$-1.303(1)$ &0.985(3) &0.600(2) &$-0.5654(2)$ \\
 2.0 & 2 & 0.2853(1) &$-1.365(1)$ &0.750(2) &0.599(2) &$-0.5196(2)$ \\
 1.7 & 2 & 0.2862(1) &$-1.416(1)$ &0.582(2) &0.595(2) &$-0.4816(2)$ \\
 1.6 & 2 & 0.2860(1) &$-1.436(1)$ &0.513(2) &0.591(2) &$-0.4661(2)$ \\
 1.2 & 3 & 0.2725(1) &$-1.510(2)$ &0.188(1) &0.537(2) &$-0.3823(2)$ \\
 1.0 & 4 & 0.2598(1) &$-1.567(2)$ &$-0.025(1)$&0.511(2) &$-0.3190(2)$ \\
 0.8 & 5 & 0.2407(1) &$-1.658(2)$ &$-0.291(1)$&0.500(2) &$-0.2290(2)$ \\
\end{tabular}
\end{center}
\end{minipage}
\end{center}
\end{table}

\begin{table}
\begin{center}
\begin{minipage}{3.2 true in}
\caption[tabthree]{ Values of the coefficients $\tau^{V_0}_i$ corresponding
  to the temporal component of the vector current $V_0$,
  similar to Table~\protect\ref{tab:one}.
\label{tab:three}}
\begin{center}
\begin{tabular}{rrrrr@{\hspace{1.2ex}}}
\multicolumn{1}{c}{$aM_0$} &
\multicolumn{1}{c}{$n$} &
\multicolumn{1}{c}{$\tau^{V_0}_0$} &
\multicolumn{1}{c}{$\tau^{V_0}_1$} &
\multicolumn{1}{c}{$\tau^{V_0}_2$} 
 \\ \hline
10.0 & 1 & 0.3549(2) &$-0.541(8) $ & $1.706(19)$\hspace{-1.2ex} \\
 7.0 & 1 & 0.3961(2) &$-0.572(6) $ & $1.339(12)$\hspace{-1.2ex} \\
 4.0 & 1 & 0.4954(2) &$-0.643(3) $ & $0.982(12)$\hspace{-1.2ex} \\
 4.0 & 2 & 0.4868(1) &$-0.695(3) $ & $0.969(6)$ \\
 3.5 & 2 & 0.5161(1) &$-0.718(3) $ & $0.907(5)$ \\
 3.0 & 2 & 0.5542(1) &$-0.752(3) $ & $0.844(4)$ \\
 2.7 & 2 & 0.5832(2) &$-0.774(2) $ & $0.815(4)$ \\
 2.5 & 2 & 0.6058(1) &$-0.794(2) $ & $0.792(3)$ \\
 2.0 & 2 & 0.6795(2) &$-0.853(2) $ & $0.744(3)$ \\
 1.7 & 2 & 0.7417(2) &$-0.899(2) $ & $0.718(2)$ \\
 1.6 & 2 & 0.7666(2) &$-0.920(2) $ & $0.707(2)$ \\
 1.2 & 3 & 0.8832(2) &$-1.034(1) $ & $0.684(1)$ \\
 1.0 & 4 & 0.9674(2) &$-1.109(1) $ & $0.682(1)$ \\
 0.8 & 5 & 1.0853(2) &$-1.202(1) $ & $0.696(1)$ \\
\end{tabular}
\end{center}
\end{minipage}
\end{center}
\end{table}

\begin{table}
\begin{center}
\begin{minipage}{5 true in}
\caption[tabfour]{Values of the coefficients $\rho^{V_k}_j$ defined
 in Eq.~(\protect\ref{cjdef}) corresponding to the spatial components of the
 vector current for various values of the bare heavy-quark mass $aM_0$ and NRQCD
  stability parameter $n$.  Uncertainties in the determinations of these
  parameters due to the use of Monte Carlo integration are included.
\label{tab:four}}
\begin{center}
\begin{tabular}{rrrrrrr}
\multicolumn{1}{c}{$aM_0$} &
\multicolumn{1}{c}{$n$} &
\multicolumn{1}{c}{$\rho^{V_k}_0$} &
\multicolumn{1}{c}{$\rho^{V_k}_1$} &
\multicolumn{1}{c}{$\rho^{V_k}_2$} &
\multicolumn{1}{c}{$\rho^{V_k}_3$} &
\multicolumn{1}{c}{$\rho^{V_k}_4$} \\ \hline
10.0 & 1 & $-0.5051(3)$ & $-0.128(5)$ & 16.968(11)\hspace{-1.2ex} & 0.991(4) & 0.069(5) \\
 7.0 & 1 & $-0.5441(2)$ & $-0.202(4)$ & 10.733(7) & 0.932(3) & 0.190(3) \\
 4.0 & 1 & $-0.5496(2)$ & $-0.346(5)$ &  4.680(7)& 0.810(7) & 0.354(7) \\
 4.0 & 2 & $-0.5744(2)$ & $-0.382(5)$ &  4.955(8)& 0.846(6) & 0.360(7) \\
 3.5 & 2 & $-0.5694(2)$ & $-0.421(5)$ &  4.011(5)& 0.816(6) & 0.392(6) \\
 3.0 & 2 & $-0.5585(2)$ & $-0.443(5)$ &  3.099(4)& 0.775(5) & 0.431(5) \\
 2.7 & 2 & $-0.5472(2)$ & $-0.470(4)$ &  2.566(4)& 0.747(5) & 0.451(5) \\
 2.5 & 2 & $-0.5366(2)$ & $-0.495(4)$ &  2.228(4)& 0.734(5) & 0.467(4) \\
 2.0 & 2 & $-0.4958(2)$ & $-0.521(3)$ &  1.415(3)& 0.678(4) & 0.505(3) \\
 1.7 & 2 & $-0.4561(3)$ & $-0.551(3)$ &  0.970(2)& 0.642(3) & 0.528(3) \\
 1.6 & 2 & $-0.4391(3)$ & $-0.564(3)$ &  0.828(2)& 0.630(3) & 0.534(2) \\
 1.2 & 3 & $-0.3679(3)$ & $-0.609(3)$ &  0.419(2)& 0.574(3) & 0.563(2) \\
 1.0 & 4 & $-0.3018(4)$ & $-0.606(3)$ &  0.254(1)& 0.532(3) & 0.570(2) \\
 0.8 & 5 & $-0.1818(5)$ & $-0.597(4)$ &  0.110(1)& 0.473(3) & 0.564(1) \\
\end{tabular}
\end{center}
\end{minipage}
\end{center}
\end{table}

\begin{table}
\begin{center}
\begin{minipage}{5 true in}
\caption[tabfour]{Values of the coefficients $\rho^{A_k}_j$
 corresponding to the spatial components of the axial-vector current,
 similar to Table~\protect\ref{tab:four}.
\label{tab:five}}
\begin{center}
\begin{tabular}{rrrrrrr}
\multicolumn{1}{c}{$aM_0$} &
\multicolumn{1}{c}{$n$} &
\multicolumn{1}{c}{$\rho^{A_k}_0$} &
\multicolumn{1}{c}{$\rho^{A_k}_1$} &
\multicolumn{1}{c}{$\rho^{A_k}_2$} &
\multicolumn{1}{c}{$\rho^{A_k}_3$} &
\multicolumn{1}{c}{$\rho^{A_k}_4$} \\ \hline
10.0 & 1 & 0.0504(2)    & $0.383(6)$ & 0.413(18)\hspace{-1.2ex} &$-0.186(6)$&$-0.1420(2)$ \\
 7.0 & 1 & $-0.0246(2)$ & $0.303(4)$ & 0.447(12)\hspace{-1.2ex} &$-0.211(4)$&$-0.0180(2)$ \\
 4.0 & 1 & $-0.1093(2)$ & $0.151(3)$ & 0.377(6)& $-0.267(2)$ & 0.1578(2) \\
 4.0 & 2 & $-0.1146(2)$ & $0.073(3)$ & 0.397(6)& $-0.166(2)$ & 0.1584(2) \\
 3.5 & 2 & $-0.1285(1)$ & $0.028(3)$ & 0.362(5)& $-0.171(2)$ & 0.1963(2) \\
 3.0 & 2 & $-0.1396(1)$ & $-0.005(3)$ & 0.337(4)&$-0.174(2)$ & 0.2375(2) \\
 2.7 & 2 & $-0.1444(2)$ & $-0.043(3)$ & 0.305(3)&$-0.174(2)$ & 0.2641(2) \\
 2.5 & 2 & $-0.1458(2)$ & $-0.070(3)$ & 0.288(3)&$-0.176(2)$ & 0.2826(2) \\
 2.0 & 2 & $-0.1401(2)$ & $-0.117(3)$ & 0.238(2)&$-0.174(2)$ & 0.3316(2) \\
 1.7 & 2 & $-0.1253(2)$ & $-0.161(3)$ & 0.200(2)&$-0.171(2)$ & 0.3625(2) \\
 1.6 & 2 & $-0.1177(2)$ & $-0.180(3)$ & 0.191(2)&$-0.166(2)$ & 0.3727(2) \\
 1.2 & 3 & $-0.0636(3)$ & $-0.275(3)$ & 0.151(1)&$-0.112(2)$ & 0.4110(2) \\
 1.0 & 4 & $-0.0058(3)$ & $-0.304(3)$ & 0.131(1)&$-0.087(2)$ & 0.4251(2) \\
 0.8 & 5 & 0.1023(5) & $-0.326(4)$ &  0.113(1)& $-0.076(2)$ & 0.4298(2) \\
\end{tabular}
\end{center}
\end{minipage}
\end{center}
\end{table}

\begin{table}
\begin{center}
\begin{minipage}{3.2 true in}
\caption[tabfive]{Values of the coefficients $\rho^{V_0}_j$
 corresponding to the temporal component of the
 vector current, similar to Table~\protect\ref{tab:four}.
\label{tab:six}}
\begin{center}
\begin{tabular}{rrrrr}
\multicolumn{1}{c}{$aM_0$} &
\multicolumn{1}{c}{$n$} &
\multicolumn{1}{c}{$\rho^{V_0}_0$} &
\multicolumn{1}{c}{$\rho^{V_0}_1$} &
\multicolumn{1}{c}{$\rho^{V_0}_2$} \\ \hline
10.0 & 1 & $0.1712(3)$ & $-0.387(9)$ & $-0.433(19)\!\!$  \\
 7.0 & 1 & $0.0599(2)$ & $-0.482(7)$ & $-0.065(12)\!\!$  \\
 4.0 & 1 & $-0.1107(3)$ & $-0.672(4)$ & $0.291(12)\!\!$ \\
 4.0 & 2 & $-0.1150(2)$ & $-0.642(5)$ &  0.305(6) \\
 3.5 & 2 & $-0.1557(2)$ & $-0.691(4)$ &  0.367(5) \\
 3.0 & 2 & $-0.2023(2)$ & $-0.725(4)$ &  0.430(4) \\
 2.7 & 2 & $-0.2341(2)$ & $-0.765(3)$ &  0.458(4) \\
 2.5 & 2 & $-0.2570(2)$ & $-0.791(3)$ &  0.481(3) \\
 2.0 & 2 & $-0.3221(2)$ & $-0.842(3)$ &  0.529(3) \\
 1.7 & 2 & $-0.3686(3)$ & $-0.891(3)$ &  0.555(2) \\
 1.6 & 2 & $-0.3861(3)$ & $-0.907(3)$ &  0.566(2) \\
 1.2 & 3 & $-0.4621(3)$ & $-0.963(3)$ &  0.589(1) \\
 1.0 & 4 & $-0.5012(4)$ & $-0.974(3)$ &  0.591(1) \\
 0.8 & 5 & $-0.5300(5)$ & $-0.994(3)$ &  0.577(1) \\
\end{tabular}
\end{center}
\end{minipage}
\end{center}
\end{table}


\begin{references}
\bibitem{errcor}
  The $\delta_{i1}\delta_{j1}$ term in Eq.~(29) of Ref.~\cite{pert1}
  is erroneous; the correct factor should be $\delta_{ij}(1-\delta_{i0})$.
\end{references}
\end{document}